\documentclass[a4paper]{jpconf}
\usepackage{graphicx}
\begin{document}

\title{Importance of axion-like particles for very-high-energy astrophysics} 

\author{Marco Roncadelli$^1$, Alessandro De Angelis$^2$ and Giorgio Galanti$^3$}

\address{$^1$ INFN, Sezione di Pavia, via A. Bassi 6, I -- 27100 Pavia, Italy}


\address{$^2$ Dipartimento di Fisica, Universit\`a di Udine, Via delle Scienze 208, I -- 33100 Udine,\\ 
and INAF and INFN, Sezioni di Trieste, Italy}



\address{$^3$ Dipartimento di Fisica, Universit\`a dell'Insubria, Via Valleggio 11, I -- 22100 Como, Italy}

\ead{marco.roncadelli@pv.infn.it}

\begin{abstract}
Several extensions of the Standard Model predict the existence of Axion-Like Particles (ALPs), very light spin-zero bosons with a two-photon coupling. ALPs can give rise to observable effects in very-high-energy astrophysics. Above roughly $100 \, {\rm GeV}$ the horizon of the observable Universe progressively shrinks as the energy increases, due to scattering of beam photons off background photons in the optical and infrared bands, which produces $e^+ e^-$ pairs. In the presence of large-scale magnetic fields photons emitted by a blazar can oscillate into ALPs on the way to us and back into photons before reaching the Earth. Since ALPs do not interact with background photons, the effective mean free path of beam photons increases, enhancing the photon survival probability. While the absorption probability  increases with energy, photon-ALP oscillations are energy-independent, and so the survival probability increases with energy compared to standard expectations. We have performed a systematic analysis of this effect, interpreting the present data on very-high-energy photons from blazars. Our predictions can be tested with presently operating Cherenkov Telescopes like H.E.S.S., MAGIC, VERITAS and CANGAROO III as well as with detectors like ARGO-YBJ and MILAGRO and with the planned Cherenkov Telescope Array and the HAWC  $\gamma$-ray observatory.  ALPs with the right properties to produce the above effects can possibly be discovered by the GammeV experiment at FERMILAB and surely by the planned photon regeneration experiment ALPS at DESY.
\end{abstract}

\section{Introduction}

In spite of the great success scored by the Standard Model of particle physics, nowadays no one regards it as the ultimate theory. Indeed, a big effort has been devoted over the last few decades to extend it in order to have a truly unified theory of all interactions including gravity. So many specific attempts have been put forward and a remarkable thing is that several of them generically predict the existence of Axion-Like Particles (ALPs), namely very light spin-zero bosons with a two-photon coupling. They closely resemble the Axion, which is the Pseudo-Goldstone boson associated to the Peccei-Quinn symmetry invented to naturally solve the strong CP problem. However, two important differences exist mainly because the Axion arises in a very specific context while in dealing with ALPs the aim is to bring out their properties in a model-independent fashion as much as 
possible~\cite{alprev}. First, only ALP-photon interaction terms are taken into account, and so ALPs are described by the Lagrangian
\begin{equation}
\label{t2230812wx}
{\cal L}_{\rm ALP} =  \frac{1}{2} \, \partial^{\mu} a \, \partial_{\mu} a - \frac{1}{2} \, m^2 \, a^2 - \, \frac{1}{4 M} \, F_{\mu\nu} \tilde{F}^{\mu\nu} a = \frac{1}{2} \, \partial^{\mu} a \, \partial_{\mu} a - \frac{1}{2} \, m^2 \, a^2 + \frac{1}{M} \, {\bf E} \cdot {\bf B} \, a~,
\end{equation}
where $a$ is the ALP field. Second, the parameters $m$ and $M$ are to be regarded as {\it unrelated} for ALPs -- while they are closely related for the Axion -- and it is merely assumed that $m \ll G_F^{- 1/2}$ and $M \gg G_F^{- 1/2}$.

Our aim is to summarize some very recent results that we have obtained concerning the implications of ALPs for very-high-energy (VHE) blazar observations. We refer to our original paper for a very thorough presentation -- which also includes cosmological effects that are discarded here for lack of space -- and for a complete list of references~\cite{dgr}.

\begin{figure}[h]
\centerline{\includegraphics[width=0.85\textwidth]{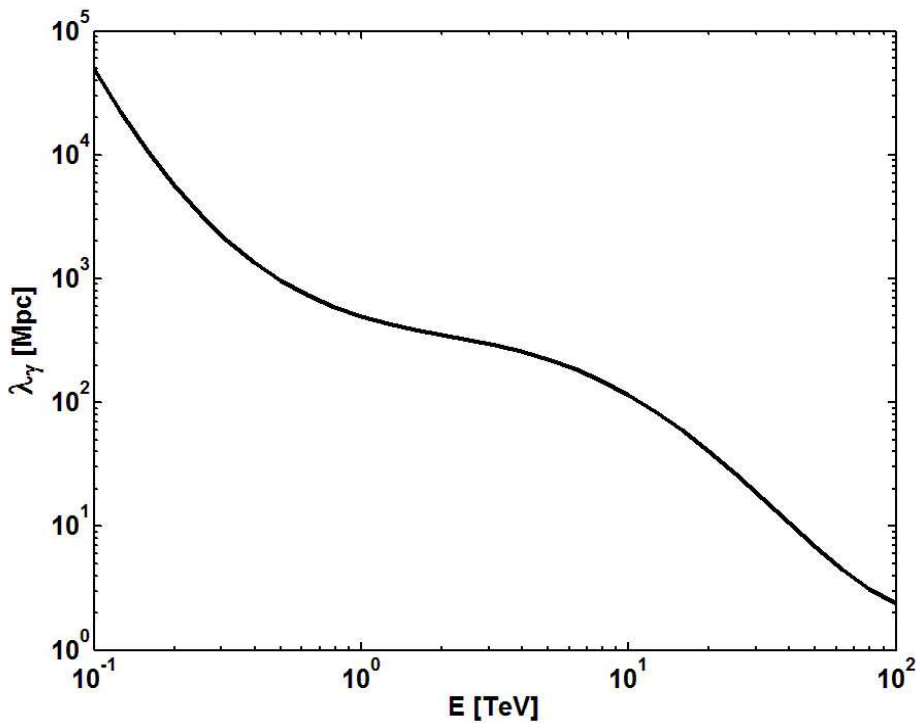}}
\caption{The mean free path of a VHE photon is plotted versus its energy within the EBL model of FRV. Only conventional physics is assumed and in particular the possibility of photon-ALP oscillations is ignored.} 
\end{figure}

\section{Extragalactic background light}

Imaging Atmospheric Cherenkov Telescopes (IACTs) have detected so far about 50 VHE blazars above $100 \, {\rm GeV}$ over distances up to the 
Gigaparsec scale and the farthest observed one is 3C279 at redshift $z = 0.536$. Given that these sources extend over a wide range of distances, not only can their intrinsic properties be inferred but also photon propagation over cosmological distances can be probed. This is particularly intriguing because VHE photons from distant sources scatter off soft background photons, thereby disappearing into $e^+ e^-$ pairs and consequently producing a dimming of the source. Since the cross-section $\sigma (\gamma \gamma \to e^+ e^-)$ peaks where the VHE photon energy $E$ and the background photon energy $\epsilon$ are related by $\epsilon \simeq (500 \, {\rm GeV}/E) \, {\rm eV}$, for VHE photons with $E > 100 \, {\rm GeV}$ the dimming is important due to the existence of the Extragalactic Background Light (EBL) in the optical and infrared bands produced by galaxies during the whole cosmic evolution. In the presence of such an energy-dependent opacity, photon propagation is controlled by the photon mean free path ${\lambda}_{\gamma}(E)$ for $\gamma \gamma \to e^+ e^-$, whose behaviour is exhibited in Figure 1 for the EBL model of Franceschini, Rodighiero and Vaccari (FRV)~\cite{frv} and shows that the horizon of the observable VHE Universe indeed rapidly shrinks above $100 \, {\rm GeV}$.

Accordingly, the observed photon spectrum $\Phi_{\rm obs}(E,D)$ is related to the emitted one $\Phi_{\rm em}(E)$ by 
\begin{equation}
\label{a1}
\Phi_{\rm obs}(E,D) = e^{- D/{\lambda}_{\gamma}(E)} \ \Phi_{\rm em}(E)~,
\end{equation}
where $D$ is the source distance. Thus, Eq.~(\ref{a1}) implies that the observed flux is {\it exponentially} suppressed both at high energies and at large distances, so that sufficiently far-away sources become hardly visible in the VHE range and their observed spectrum should be {\it much steeper} than the emitted one.

\section{DARMA scenario}

Our proposal -- to be referred to as the DARMA scenario, such an acronym coming from the initials of the authors of the first article on the subject~\cite{drm} -- can be sketched as follows~\cite{dgr,drm}. 

Since ALPs are characterized by a coupling to two photons, in the presence of an external magnetic field $\bf B$ the interaction eigenstates differ from the propagation eigenstates, so that photon-ALP oscillations show up (much in the same way as it happens for massive neutrinos). Photons are supposed to be emitted by a blazar in the usual way. In the presence of extragalactic magnetic fields some of them can turn into ALPs. Further, some of the produced ALPs can convert back into photons and ultimately be detected. In free space this would obviously produce a flux dimming. Remarkably enough, because of the EBL such a double conversion can make the observed flux {\it considerably larger} than in the standard situation. This is due to the fact that ALPs do {\it not} undergo EBL absorption. As a consequence, the observed photons travel a distance in excess of ${\lambda}_{\gamma}(E)$ and Eq. (\ref{a1}) becomes
\begin{equation}
\label{a1bis}
\Phi_{\rm obs}(E,D) = e^{- D/{\lambda}_{\gamma , {\rm eff}}(E)} \ \Phi_{\rm em}(E)~,
\end{equation}
which shows that even a {\it small} increase of the photon effective mean free path ${\lambda}_{\gamma , {\rm eff}}(E)$ gives rise to a {\it large} enhancement of the observed flux. It turns out that the DARMA mechanism makes ${\lambda}_{\gamma , {\rm eff}}(E)$ shallower than ${\lambda}_{\gamma}(E)$ although it remains a decreasing function of $E$. So, the resulting observed spectrum is {\it much harder} than the one predicted by Eq. (\ref{a1}), thereby ensuring agreement with observations even for a standard emission spectrum. 

\section{Predicted energy spectra}

We suppose that the large-scale magnetic fields ${\bf B}$ are fairly strong but still consistent both with Auger observations and the presently available 
upper bound $B < 6 \, {\rm nG}$. Unfortunately, their morphology is largely unknown and so it is usually supposed that they have a domain-like structure with domain size in the range $1\, {\rm Mpc} < L_{\rm dom} < 10 \, {\rm Mpc}$. We also take $M > 10^{11} \, {\rm GeV}$, in agreement with all available bounds. Since all observable effects depend on $B/M$, we introduce the parameter $\xi \equiv (B/{\rm nG})(10^{11} \, {\rm GeV}/M)$. Then the above constraints translate into $\xi < 6$.

\begin{figure}
\centerline{\includegraphics[width=1.05\textwidth]{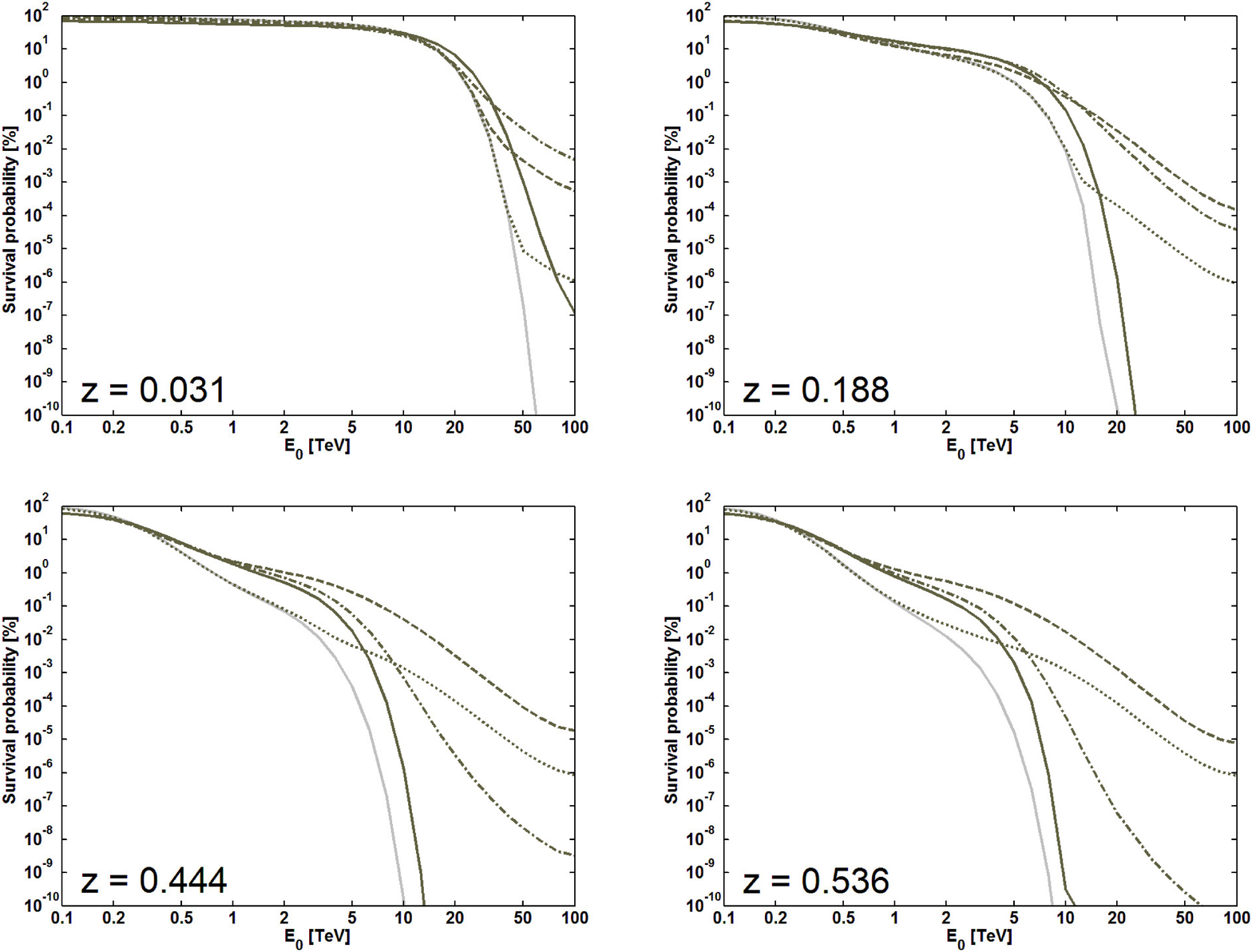}}
\caption{Behaviour of $P^{\rm DARMA}_{\gamma \to \gamma}$ versus the observed energy $E_0$ for four different redshifts $z=0.031$, $z=0.188$, $z=0.444$ and $z=0.536$. The solid black line corresponds to $\xi =5.0$, the dotted-dashed line to $\xi =1.0$, the dashed line to $\xi =0.5$, the dotted line to $\xi =0.1$ and the solid grey line to conventional physics. We have taken $L_{\rm dom} = 4 \, {\rm Mpc}$.} \label{Fig:MV}
\end{figure}

We evaluate the photon survival probability $P_{\gamma \to \gamma}(E,D)$ when allowance is made for photon-ALP oscillations as well as for photon absorption from the EBL, so that Eq. (\ref{a1bis}) gets replaced by
\begin{equation}
\label{a0as}
\Phi_{\rm obs}(E,D) = P_{\gamma \to \gamma}^{\rm DARMA} (E,D) \, \Phi_{\rm em}(E)~. 
\end{equation}

Our procedure is as follows. We first solve exactly the beam propagation equation over a single magnetic domain, assuming that the EBL is described by the FRV model. Starting with an unpolarized photon beam, we next propagate it by iterating the single-domain solution as many times as the number of domains crossed by the beam, taking each time a {\it random} value for the angle between ${\bf B}$ and a fixed overall fiducial direction. We repeat such a procedure $5000$ times and finally we average over all these realizations of the propagation process. The resulting photon survival probabilities are plotted in Figure 2 for various values of the free parameters.  We see that a ``boost factor'' (i.e., an enhancement with respect to standard physics) as large as 10 is achieved at energies of a few TeV for farther sources, at redshift values of about 0.2 and larger.


\section*{References}
\begin{thebibliography}{99}
\bibitem{alprev} Jaeckel J, Ringwald A, 2010 {\it Ann. Rev. Nucl. Part. Sci.} {\bf 60} 405 
\bibitem{dgr} De Angelis A, Galanti G, Roncadelli M 2011 {\it Phys. Rev.} D {\bf 84} 105030
\bibitem{frv} Franceschini A, Rodighiero G, Vaccari M 2008 {\it Astron. Astrophys.} {\bf 487} 837 
\bibitem{drm} De Angelis A, Roncadelli M and Mansutti O 2007 {\it Phys. Rev.} D {\bf 76} 121301 

\end{thebibliography}
\end{document}